\documentclass[american,english,manuscript]{aastex}
\usepackage[T1]{fontenc}
\usepackage[latin9]{inputenc}
\usepackage{bm}
\usepackage{amsbsy}
\usepackage{amssymb}
\usepackage{graphicx}
\usepackage{esint}

\makeatletter
\pdfoutput=1

\usepackage{bm}\usepackage{amsbsy}

\usepackage{bm}

\newcommand{\mathe}{\mathrm{e}}
\newcommand{\mathd}{\mathrm{d}}
\shorttitle{Turbulent cross-helicity in mean-field dynamo}
\shortauthors{Pipin et al.}
\bibpunct[,]{(}{)}{;}{a}{,}{,}

\makeatother

\usepackage{babel}
\begin{document}

\title{Turbulent cross-helicity in the mean-field solar dynamo problem}

\author{V.V. Pipin$^{1-4}$, K.M. Kuzanyan$^{1,5}$, H.Zhang$^{1}$ and A.G.
Kosovichev$^{4}$}

\affil{$^{1}$National Astronomical Observatories, Chinese Academy of Sciences,
Beijing 100012, China:pip@iszf.irk.ru \\
 $^{2}$Institute of Solar-Terrestrial Physics, Russian Academy of
Sciences, \\
 $^{3}$ Institute of Geophysics and Planetary Physics, UCLA, Los
Angeles, CA 90065, USA \\
 $^{4}$Hansen Experimental Physics Laboratory, Stanford University,
Stanford, CA 94305, USA \\
 $^{5}$Institute for Solar-Terrestrial Magnetism \& Ionosphere \&
Radiowave Propagation, Russian Academy of Sciences }
\begin{abstract}
We study the dynamical and statistical properties of turbulent cross-helicity
(correlation of the aligned fluctuating velocity and magnetic field
components). We derive an equation governing generation and evolution
of the turbulent cross-helicity and discuss its meaning for the dynamo.
Using symmetry properties of the problem we suggest a general expression
for the turbulent cross-helicity. Effects of the density stratification,
large-scale magnetic fields, differential rotation and turbulent convection
are taken into account. We investigate the relative contribution of
these effects to the cross-helicity evolution for two kinds of dynamo
models of the solar cycle: a distributed mean-field model
and a flux-transport dynamo model. We show that the contribution from
the density stratification follows the evolution of the radial magnetic
field, while large-scale electric currents produce a more complicated
pattern of the cross-helicity of the comparable magnitude.
The pattern
of the cross-helicity evolution strongly depends on details of the
dynamo mechanism. Thus, we anticipate that direct observations of
the cross-helicity on the Sun may serve for the diagnostic purpose
of the solar dynamo process. 
\end{abstract}


\section{Introduction}

The cross-helicity conservation law has been established by \citet{wolt}
and \citet{moff}. It states that in perfectly conductive media the
parallel components of magnetic and velocity fields do not interact.
{ More precisely, the conservation law states that if the velocity field
of plasma, $\mathbf{U}$, and the magnetic induction vector field,
$\mathbf{B}$, are confined in the volume $V$ then the integral $I=\int_{V}\mathbf{U}\cdot\mathbf{B}dV$
is time-invariant if the dissipative processes are absent.}

The {}``Alfvénic'' state of magnetohydrodynamical (MHD) turbulence,
in which the fluctuating magnetic and velocity fields are aligned,
is commonly considered as a preferred state in MHD relaxation processes,
\citep{wolt,2008PhRvL.100h5003M}. Such state is also accompanied
by quenching of the turbulent magnetic field generation effects (e.g.,
the $\alpha$-effect) due to the back reaction of large-scale magnetic
fields on the turbulent motions. Therefore, the cross-helicity can
be a useful quantity for diagnostics of non-linear turbulent dynamo
processes in the solar convection zone, (see, e.g., \citealp{2003A&A...409.1097K}).
\citet{1990PhFlB...2.1589Y} and \citet{1999PhFl...11.2307Y} considered
the cross-helicity as a part of the dynamo mechanism in turbulent
astrophysical flows. The role of the cross-helicity conservation for
this type of dynamo was explored recently by \citet{2009MNRAS.399..273S}.
The results of \citet{2010SoPh..tmp..241R} suggested that the cross-helicity
parameter that could be observed on the solar surface is a manifestation
of solely near-surface physical processes. On the other hand, the
results of numerical simulations \citep{2006PhRvL..97y5002M,2009ApJ...699L..39B}
showed the existence of rapid local alignments of turbulent velocity
and magnetic field in the presence of spatially nonuniform pressure
and kinetic energy of turbulent fluctuations. It is notable that these
locally aligned turbulent velocity and magnetic field structures spatially
dominate even in the case when the mean cross-helicity is zero. Some
initial attempts to measure the cross-helicity from SOHO/MDI data
\citep{1995SoPh..162..129S} were made by \citet{2011SoPh..tmp...39Z}.

In observations of the cross-helicity at the solar surface the processes
of the {}``local'' and {}``global'' alignments may be mixed. To
isolate the effect due to the local alignment we have to choose appropriate
spatial and temporal scales for averaging. In this paper, we show
that the large-scale distribution of the averaged cross-helicity can
be affected by the solar dynamo operating in the deep convection zone,
although the scale separation problem remains a difficult issue of
the theory.

The theoretical approach to study the solar dynamo is based on the
mean-field magnetohydrodynamics \citep{krarad80}. In the following
part of the paper we propose  a general expression
for the turbulent cross-helicity by the use of the transformation 
symmetry properties. Then we estimate the coefficients in this
expression
 by means of the $\tau$- approximation in the turbulence
theory \citep{1970JFM....41..363O,vain-kit:83,black-bran:02,2002ApJ...572..685F,2002PhRvL..89z5007B,2005PhR...417....1B}.
Finally, we illustrate the major contributions to the cross-helicity
evolution using two dynamo models of the solar magnetic cycle.

\section{Transformation symmetry properties and  cross-helicity}

Following \citet{krarad80}, we split the physical quantities of the
turbulent conducting fluid into the mean and randomly fluctuating
parts. The mean parts are defined as \textit{ensemble} averages of
the corresponding properties. Magnetic field $\mathbf{B}$ and velocity
$\mathbf{U}$ are decomposed as follows: 
\begin{equation}
\mathbf{B}=\overline{\mathbf{B}}+\mathbf{b},\,\mathbf{U}=\overline{\mathbf{U}}+\mathbf{u}.\label{eq:decom}
\end{equation}
Hereafter, we use capital letters with a bar above for the mean-field
(large-scale) properties and the small letters for the fluctuating
(small-scale) parts. As shown by \citet{krarad80} and \citet{1980AN....301..101R}
the general structure of the mean electromotive force of turbulent
flows, $\mathbf{\mathcal{E}}\equiv\overline{\mathbf{u}\times\mathbf{b}}$,
can be reconstructed by using the transformation symmetry properties
of the basic physical quantities and their products involved in the
problem. We deal with the cross-helicity pseudo-scalar $\overline{\mathbf{u}\cdot\mathbf{b}}$
in the similar way. 

Any pseudo-scalar can be expressed either as a
tensor product of a tensor with pseudo-tensor, or as a scalar product
of vector and pseudo-vector.
General expression for $\overline{\mathbf{u}\cdot\mathbf{b}}$ may
have a fairly complicated structure. We restrict ourselves in studying
the effects that can be important for the solar convection zone dynamics
and solar dynamo. Therefore, we have to take into account the stratification
of the thermodynamic and turbulent parameters. Their contribution
to the solar dynamics is given by { gradients} of the turbulent
kinetic fluctuations and magnetic energy density, e.g., $\boldsymbol{\nabla}\left(\overline{\rho}\overline{\mathbf{u}^{2}}\right)$
and $\boldsymbol{\nabla}\left({\displaystyle \frac{\overline{\mathbf{b}^{2}}}{2\mu_{0}}}\right)$
(all quantities are ordinary vectors). { Note, that the mean density
is sufficient to describe the effects of the thermodynamic stratification
in the polytropic atmosphere. This assumption will be used hereafter
 in the paper.}
Also, we have to take into account effects of the global rotation
(pseudo-vector $\boldsymbol{\Omega}$) and large-scale shear,
 $\nabla_{i}\overline{U}_{j}$
(tensor). { This tensor can be decomposed
as follows: $\nabla_{i}\overline{U}_{j}=\frac{1}{2}\varepsilon_{ijp}W_{p}+\left(\boldsymbol{\nabla}\overline{\mathbf{U}}\right)_{\{i,j\}}$,
where, $\mathbf{W}=\left(\nabla\times\mathbf{\overline{U}}\right)$,
(pseudo-vector) and a strain tensor, $\left(\boldsymbol{\nabla}\overline{\mathbf{U}}\right)_{\{i,j\}}=\frac{1}{2}\left(\nabla_{i}\overline{U}_{j}+\nabla_{j}\overline{U}_{i}\right)$.}
We assume that $\overline{\mathbf{u}\cdot\mathbf{b}}=0$ in the absence
of large-scale magnetic field and take into account
effects of the large-scale magnetic field, $\overline{\mathbf{B}}$
(pseudo-vector), and its spatial derivatives, $\nabla_{i}\overline{B}_{j}$.
Similar to the large-scale shear flows we decompose the contribution
of the non-uniform magnetic field into effects of large-scale electric
current $\overline{\mathbf{J}}=\boldsymbol{\nabla}\times\overline{\mathbf{B}}/\mu_{0}$
(vector) and magnetic strain tensor, $\left(\boldsymbol{\nabla}\overline{\mathbf{B}}\right)_{\{i,j\}}=\frac{1}{2}\left(\nabla_{i}\overline{B}_{j}+\nabla_{j}\overline{B}_{i}\right)$
( pseudo-tensor). { By analogy with \citet{1980AN....301..101R},  and also assuming
the scale separation $\{\ell,\tau_{c}\}\ll\{L,T\}$ of the typical
small-scale $(\ell,\tau_{c})$ and the large-scale $(L,T)$ spatial
and temporal variations of the velocity and magnetic fields,  we write
a combination of above effects as follows:}
\begin{eqnarray}
\overline{\mathbf{u}\cdot\mathbf{b}} & = & \frac{\tau_{c}}{\overline{\rho}}\left(\overline{\mathbf{B}}\cdot\nabla\right)\left(\kappa_{1}\bar{\rho}\overline{\mathbf{u}^{2}}+\kappa_{2}\frac{\overline{\mathbf{b}^{2}}}{2\mu_{0}}\right)\label{eq:symmetry}\\
 & + & \kappa_{3}\mu_{0}\overline{\mathbf{W}}\cdot\overline{\mathbf{J}}+\kappa_{4}\mu_{0}\boldsymbol{\Omega}\cdot\overline{\mathbf{J}}+\kappa_{5}\left(\boldsymbol{\nabla}\overline{\mathbf{U}}\right)_{\{i,j\}}\left(\boldsymbol{\nabla}\overline{\mathbf{B}}\right)_{\{i,j\}}+o\left(\frac{\ell}{L}\right),\nonumber 
\end{eqnarray}
{ where, $\kappa_{1-5}$ are coefficients that needs to be determined).
Having in mind the transformation symmetry properties of the quantities
that are involved in the problem, we, of course, can construct the
pseudo-scalar $\overline{\mathbf{u}\cdot\mathbf{b}}$ in many other
ways. In principle, the Eq. (\ref{eq:symmetry}) may include the terms
like, $\mathbf{G\cdot\boldsymbol{\Omega}},\,\mathbf{\overline{U}}\cdot\boldsymbol{\Omega}$
(and similar), or the terms like, $\mathbf{\overline{U}\cdot\overline{B}},\,\overline{\mathbf{B}}\cdot\mathbf{\overline{J}}$
(and similar), or even higher order combination of the physical quantities.
Later, we will see that such terms do not appear in our analysis of
the momentum and induction equations of the fluctuating velocity and
magnetic fields if we restrict ourselves to the case of the weak large-scale
magnetic field $\mathbf{\overline{B}}$. In this sense the Equation
(\ref{eq:symmetry}) may be incomplete and should be considered with
cautions in analysis of the observational data.}


The physical interpretation of the first term in Eq. (\ref{eq:symmetry})
was given by \citet{2010SoPh..tmp..241R}. Consider the turbulent
medium permeated by the large-scale field $\mathbf{\overline{B}}$.
Convective elements rising, with velocity $\mathbf{u}$ expand, $\left(\nabla\cdot\mathbf{u}\right)>0$,
and induce a fluctuating magnetic field, $\mathbf{b}\approx-\tau_{c}\mathbf{\overline{B}}\left(\nabla\cdot\mathbf{u}\right)$
(sign is anti-correlated with sign of $\left(\nabla\cdot\mathbf{u}\right)$).
The same is valid for descending and contracting convective elements.
In the anelastic approximation \citep{1969JAtS...26..448G}, $\nabla\cdot\overline{\rho}\mathbf{u}\equiv0$
and $\left(\nabla\cdot\mathbf{u}\right)=-\left(\mathbf{u}\cdot\boldsymbol{\nabla}\log\overline{\rho}\right)$.
Therefore the sign of $\overline{\mathbf{u}\cdot\mathbf{b}}$ is opposite
to the sign of $\overline{\mathbf{B}}$ ({ because} ${\nabla_r}\log\overline{\rho}<0$).
The other terms in Eq. (\ref{eq:symmetry}) are less easy to interpret.
{ The effects
of density fluctuations are not taken into account in this
consideration.
For weakly compressible (subsonic) convective flows}, ${\displaystyle \frac{\overline{\mathbf{u}^{2}}}{C_{S}^{2}}\ll1}$
($C_{s}$ is the sound speed), the contribution of the buoyancy forces
to the cross-helicity is proportional to $\sim\kappa_{6}{\displaystyle \frac{\overline{\mathbf{u}^{2}}}{C_{S}^{2}}\tau_{c}\left(\mathbf{g}\cdot\mathbf{\overline{B}}\right)}$,
where $\mathbf{g}$ is the gravity acceleration. This effect may be
of the same order magnitude as the first term in Eq. (\ref{eq:symmetry}),
particularly close to the surface where ${\displaystyle \frac{\overline{\mathbf{u}^{2}}}{C_{S}^{2}}}$
can be significant. However, we do not consider the density fluctuations
effects, and limit our study to the anelastic approximation, which
allows us to investigate the turbulent cross-helicity analytically.
In what follows we estimate the one-point cross-helicity correlations using
two different approaches in the mean-field theory, based on the so-called
$\tau$- approximation of the turbulence theory, in which third-order
correlations are approximated by a dissipative term with a characteristic
dissipation time $\tau_{c}$
\citep{1970JFM....41..363O,vain-kit:83,1996A&A...307..293K,
2000PhRvE..61.5202R,2002PhRvL..89z5007B,black-bran:02,2005PhR...417....1B}.
Using the $\tau$- approximation we deal with small deviations in the
turbulent state resulting from large-scale magnetic fields and flows.
In this approximation it is assumed that the background turbulent
state (in the absence of the large-scale fields) has a priori defined
statistical properties. Our first approach is based on the cross-helicity
evolution equation and averaging turbulent properties in the physical
coordinate space. It is simpler but less accurate than the second
approach that uses a two-scale Fourier transformation for equations
governing the evolution of fluctuating velocity and magnetic fields.

\section{Mean-field theory of the cross-helicity}

\subsection{Calculation of $\overline{\mathbf{u}\cdot\mathbf{b}}$ by averaging
in the physical coordinate space}

As pointed out by \citet{2000ApJ...537.1039Y} the evolution equation
for the cross-helicity is useful for exploring mean-field dynamo properties
based on the cross-helicity effects (also see, \citealp{2009MNRAS.399..273S}).
We start with the  ideal MHD induction and momentum equations written in a coordinate frame rotating with a constant
angular velocity $\boldsymbol{\Omega}$ : 
\begin{eqnarray}
\frac{\partial\mathbf{B}}{\partial t} & = & \nabla\times\left(\mathbf{U}\times\mathbf{B}\right)\,,\label{eq:induc}\\
\rho\partial_{t}\mathbf{U}+\rho\left(\mathbf{U\cdot}\nabla\right)\mathbf{U}+2\rho\left(\boldsymbol{\Omega}\times\mathbf{U}\right)
& = &
\frac{\left(\nabla\times\mathbf{B}\right)\times\mathbf{B}}{\mu_{0}}-\rho\mathbf{\nabla}\Psi_{g}-\mathbf{\nabla}p
+\rho\mathbf{f}'\,.\label{eq:moment}
\end{eqnarray}
Here, $\mathbf{U}$ and $\mathbf{B}$ are the velocity and magnetic
field vectors, $p$ is the gas pressure, $\Psi_{g}$ is the gravitational
potential and $\mathbf{f}'$  is the external random force. Calculating the scalar products of the first equation with
$\mathbf{U}$, and the second equation with $\mathbf{B}/\rho$, and
summing them up, we arrive to: 
\begin{eqnarray}
\partial_{t}\left(\mathbf{U\cdot
    B}\right)+2\boldsymbol{\Omega}\cdot\left(\mathbf{U}\times\mathbf{B}\right)\!\!
& = &\!\! -\boldsymbol{\nabla}\cdot\left\{
  \mathbf{U}\left(\mathbf{U}\cdot\mathbf{B}\right)+\mathbf{B}\left(\Psi_{g}
+\frac{p}{\rho}\frac{\gamma}{\gamma-1}-\frac{\mathbf{U}^{2}}{2}\right)\right\}
+ \left(\mathbf{B}\cdot\mathbf{f}'\right)
 \label{eq:cross-hel-C}
\end{eqnarray}
Here, we assumed the polytropic pressure law, $p\sim\rho^{\gamma}$
(this gives
$\displaystyle{\frac{\mathbf{\nabla}p}{\rho}=\mathbf{\nabla}\left(\frac{\gamma}{\gamma-1}\frac{p}{\rho}\right)}$).
{ Eq.(\ref{eq:cross-hel-C}) describes the cross-helicity
conservation of barotrophic fluids (Woltjer, 1958)
 in the inertial frame (case $\boldsymbol{\Omega}=0$ )
 and for a potential external force,
 $\mathbf{f}'=-\nabla\phi$.
 The conservation law states that} $\partial_{t}\int_{V}\mathbf{U}\cdot\mathbf{B}dV$=0
in a volume $V$ if the normal components of ${\bf U}$ and ${\bf B}$
vanish at the boundary. In the turbulent media the cross-helicity
is not conserved because nothing prevents it to cascade from large
scales to small scales where it is dissipated. This problem was recently
discussed by \citet{2010SoPh..tmp..241R}.\textbf{ }

{ As discussed above, in our study we neglect the density fluctuations
effects and adopt the anelastic approximation,}
$\mbox{{\rm div}}\bar{\rho}\bar{\bm{U}}=\mbox{{\rm div}}\bar{\rho}{\bm{u}}=0$,
\citep{1969JAtS...26..448G}. Substituting Eqs.(\ref{eq:decom}) in
Eqs.(\ref{eq:induc}) and Eq.(\ref{eq:moment}) and { averaging them
over an ensemble of fluctuations} we get 
\begin{eqnarray}
\partial_{t}\overline{\mathbf{B}} & = & \mathbf{\nabla}\times\left(\mathbf{\mathcal{E}}+\overline{\mathbf{U}}\times\overline{\mathbf{B}}\right),\:\mathbf{\mathcal{E}}=\overline{\mathbf{u}\times\mathbf{b}}\\
\overline{\rho}\partial_{t}\overline{\mathbf{U}}+2\overline{\rho}\left(\boldsymbol{\Omega}\times\overline{\mathbf{U}}\right) & = & \frac{\left(\nabla\times\overline{\mathbf{B}}\right)\times\mathbf{\overline{B}}}{\mu_{0}}-\overline{\rho}\left(\mathbf{\overline{U}\cdot}\nabla\right)\mathbf{\overline{U}}-\overline{\rho}\mathbf{\nabla}\Psi_{g}\\
 & - & \mathbf{\nabla}\overline{p}+\frac{\overline{\left(\nabla\times\mathbf{b}\right)\times\mathbf{b}}}{\mu_{0}}-\overline{\rho}\overline{\left(\mathbf{u}\mathbf{\cdot}\nabla\right)\mathbf{u}}.\nonumber 
\end{eqnarray}
Similarly we derive the following evolution equation for $\overline{\mathbf{u\cdot}\mathbf{b}}$
after subtracting the equation for the mean-field cross-helicity $\overline{\mathbf{U}}\cdot\overline{\mathbf{B}}$
from Eq.(\ref{eq:cross-hel-C}): 
\begin{eqnarray}
\partial_{t}\left(\overline{\mathbf{u\cdot}\mathbf{b}}\right) & = & -\mathbf{\nabla\cdot}\overline{\mathcal{F}}^{C}-\mathbf{\mathcal{E}}\cdot\left(2\boldsymbol{\Omega}+\left(\nabla\times\mathbf{\overline{U}}\right)\right)+\frac{\overline{B}_{i}}{\overline{\rho}}\nabla_{j}T_{ij}-\frac{\left(\overline{\mathbf{u\cdot}\mathbf{b}}\right)}{\tau_{c}}\label{eq: cross-hel-tur}\\
T_{ij} & = & \overline{\rho}\overline{u_{i}u_{j}}-\frac{1}{\mu_{0}}\left(\overline{b_{i}b_{j}}-\delta_{ij}\frac{\overline{\mathbf{b}^{2}}}{2}\right),\\
\overline{\mathcal{F}}^{C} & = &
\mathbf{\overline{U}}\,\overline{\mathbf{u\cdot}\mathbf{b}}
+\overline{B}_{i}\overline{{u_{i}}\mathbf{u}}-\mathbf{\overline{B}}\frac{\overline{\mathbf{u}^{2}}}{2}.\label{eq:turb-stres0}
\end{eqnarray}
{ Here, following the $\tau$-approximation approach
\citep{vain-kit:83,1996A&A...307..293K,
2002PhRvL..89z5007B,black-bran:02} in the Eq.(\ref{eq:cross-hel-C}), we
replace the third-order correlations of the fluctuating parameters, and
the terms $\overline{\mathbf{b}\cdot\mathbf{f}'}$
with a relaxation term
${\displaystyle -\frac{\left(\overline{\mathbf{u\cdot}\mathbf{b}}\right)}{\tau_{c}}}$,
here $\tau_{c}$ is a typical time scale of the turbulent motions.
Approximating these complicated contributions with the
simple 
term ${\displaystyle
  -\frac{\left(\overline{\mathbf{u\cdot}\mathbf{b}}\right)}{\tau_{c}}}$
  has to be considered as a questionable assumption. It involves
  additional assumptions (see \citealp{2007GApFD.101..117R}), e.g., it is assumed that the second-order
correlations in Eq.(8) do not vary 
significantly on the time scale of  $\tau_{c}$. This assumption
is consistent with  scale separation between the
mean and fluctuating quantities in the mean-field
magnetohydrodynamics.
 The reader can find a comprehensive discussion
 of the $\tau$ -approximation in the above cited papers.}

In our study we discard the cross-helicity flux $\overline{\mathcal{F}}^{C}$
in the further analysis. This can be partly justified if we average
over a sufficiently large volume of the fluid in which the vector
field $\overline{\mathcal{F}}^{C}$ is confined. For example, if the averaging is carried out over
an ensemble of the turbulent fields, then we assume that this ensemble
is sufficiently representative, and that all members of the ensemble
are confined and uniformly distributed in the volume. 

In addition, in
the turbulent stress tensor $T_{ij}$ we consider only the contribution
of the turbulent kinetic and magnetic pressure as formulated in Eq.(\ref{eq:symmetry}):
\begin{eqnarray}
T_{ij} & \approx & \delta_{ij}\left(\kappa_{1}\bar{\rho}\overline{\mathbf{u}^{2}}+\kappa_{2}\frac{\overline{\mathbf{b}^{2}}}{2\mu_{0}}\right),\label{eq:turb-stresses}
\end{eqnarray}
In a linear approximation (the weak mean-field case), $\kappa_{1,2}=1/3$
\citep{1996A&A...307..293K}. In general, the turbulent stress tensor
contains terms governing the differential rotation and the meridional
circulation (\citealp{1995A&A...299..446K}).
We assume that these terms are stationary and do not contribute to
the cross-helicity evolution. Therefore, we simplify Eq.(\ref{eq: cross-hel-tur})
as: 
\begin{eqnarray}
\partial_{t}\left(\overline{\mathbf{u\cdot}\mathbf{b}}\right) & = & -\mathbf{\mathcal{E}}\cdot\left(2\boldsymbol{\Omega}+\mathbf{W}\right)+\frac{1}{3\overline{\rho}}\left(\overline{\mathbf{B}}\cdot\nabla\right)\left(\bar{\rho}\overline{\mathbf{u}^{2}}+\frac{\overline{\mathbf{b}^{2}}}{2\mu_{0}}\right)-\frac{\left(\overline{\mathbf{u\cdot}\mathbf{b}}\right)}{\tau_{c}},\label{eq:tur-cross-evol}
\end{eqnarray}
Thus, the major sources of the turbulent cross-helicity are due to
the mean electromotive force, the mean vorticity, $2\boldsymbol{\Omega}+\mathbf{\overline{W}}$,
and the gradients of the turbulent energy. { For the simplest representation
of the mean electromotive force}, $\mathbf{\mathcal{E}}\approx\alpha\overline{\mathbf{B}}-\eta_{T}\mathbf{\nabla}\times\overline{\mathbf{B}}+{\displaystyle o\left(\frac{\ell}{L}\right)}$,
where the first term represents $\alpha$-effect (here, $\alpha$
is pseudo-scalar) and $\eta_{T}$ is the turbulent diffusion coefficient,
we can write 
\begin{eqnarray}
\partial_{t}\left(\overline{\mathbf{u\cdot}\mathbf{b}}\right) & = & \frac{1}{3\overline{\rho}}\left(\overline{\mathbf{B}}\cdot\nabla\right)\left(\bar{\rho}\overline{\mathbf{u}^{2}}+\frac{\overline{\mathbf{b}^{2}}}{2\mu_{0}}\right)\label{eq:tur-cross-ev0}\\
 & - & \alpha\left(\overline{\mathbf{B}}\cdot\left(2\boldsymbol{\Omega}+\overline{\mathbf{W}}\right)\right)+\mu_{0}\eta_{T}\left(2\boldsymbol{\Omega}+\overline{\mathbf{W}}\right)\cdot\overline{\mathbf{J}}-\frac{\overline{\mathbf{u\cdot}\mathbf{b}}}{\tau_{c}},\nonumber 
\end{eqnarray}
The first two terms of this equation show that the mean cross-helicity
is generated in the presence of the large-scale magnetic field that
permeates the stratified turbulent medium. Another important contribution
represented by the third term is due to the large-scale electric current
and the mean vorticity. A similar equation was used by \citet{2000ApJ...537.1039Y}
for discussing a mean-field dynamo scenario based on the cross-helicity
effects. 

{ A complete expression for the mean  electromotive
force was used by \cite{ku2007} who investigated the time-evolution of
the cross-helicity produced by the mean-field dynamo in the solar
convection zone. Their results showed that the cross-helicity sign
alternates  during the solar cycle with the amplitude of about
$\pm 2 \mathrm{G\ km S^{-1}}$. The generated patterns of the 
cross-helicity  depend on the  structure of the mean
electromotive force which is employed in the dynamo model. Their
model, however does not include the stratification effects associated with 
the second term in Eq.(\ref{eq:tur-cross-evol}).}

To study the distributions of the cross-helicity on time intervals
much shorter than the period of the solar cycle, we can neglect the
time derivative in Eq.(\ref{eq:tur-cross-ev0}) and find:
\begin{equation}
\overline{\mathbf{u\cdot}\mathbf{b}}\approx\frac{\tau_{c}}{3\overline{\rho}}\left(\overline{\mathbf{B}}\cdot\nabla\right)\left(\bar{\rho}\overline{\mathbf{u}^{2}}+\frac{\overline{\mathbf{b}^{2}}}{2\mu_{0}}\right)-\alpha\tau_{c}\left(\overline{\mathbf{B}}\cdot\left(2\boldsymbol{\Omega}+\overline{\mathbf{W}}\right)\right)+\mu_{0}\eta_{T}\tau_{c}\left(2\boldsymbol{\Omega}+\overline{\mathbf{W}}\right)\cdot\overline{\mathbf{J}}\label{eq:crh-sta}
\end{equation}
To estimate the magnitude of the cross-helicity in the near surface
layers we use additional assumptions. First, we assume that the energy
of the fluctuating magnetic fields can be expressed through the kinetic
energy of convective motions, ${\displaystyle \bar{\frac{\overline{\mathbf{b}^{2}}}{2\mu_{0}}=\varepsilon\overline{\rho}}\overline{\mathbf{u}^{2}}}$,
and $\varepsilon=1$ is the energy equipartition condition. The convection
in the surface layers is short lived in comparison to the period of
the global rotation. For these condition the $\alpha$-effect can
be estimated as following $\alpha\approx\eta_{T}\tau_{c}\mbox{\ensuremath{\Omega}}\left(\varepsilon+1\right)\nabla\log\mbox{\ensuremath{\left(\bar{\rho}\overline{u^{2}}\right)}}$(see,
\citealp{pi08Gafd}). Then, the second term in Eq.(\ref{eq:crh-sta})
is much smaller than the first term because of $\tau_{c}\mbox{\ensuremath{\Omega}}\ll1$
in the subsurface layer. Therefore we will neglect this contribution
in the further study. For the further comparison we introduce the
density stratification scale parameter, $\mathbf{G}\equiv\nabla\log\bar{\rho}$,
and rewrite Eq.(\ref{eq:crh-sta}) as follows
\begin{equation}
\overline{\mathbf{u\cdot}\mathbf{b}}\approx\eta_{T}\left(\varepsilon+1\right)\left\{ \left(\overline{\mathbf{B}}\cdot\mathbf{G}\right)+\left(\mathbf{\overline{B}}\cdot\mathbf{\mathbf{\nabla}}\right)\log\left(\overline{\mathbf{u}^{2}}\right)\right\} +\mu_{0}\eta_{T}\tau_{c}\left(2\boldsymbol{\Omega}+\overline{\mathbf{W}}\right)\cdot\overline{\mathbf{J}}\label{eq:crh-sta-1}
\end{equation}
If we neglect the effect of fluctuating magnetic field, we find that
the contribution of density stratification agrees with \citep{2010SoPh..tmp..241R},
but the contribution of the turbulent intensity stratification
is larger by a factor of 2, in our case. The difference can be explained
by the approximations made in calculation of Eq.(\ref{eq:crh-sta-1}).
In particular, the contributions of $\left(\boldsymbol{\nabla}\overline{\mathbf{U}}\right)_{\{i,j\}}\left(\boldsymbol{\nabla}\overline{\mathbf{B}}\right)_{\{i,j\}}$
are not included in Eq.(\ref{eq:crh-sta-1}). The main purpose of
this approach is to demonstrate the basic physical contributions to
the mean cross-helicity in the solar conditions. We have to point
out that in the derivation of Eq.(\ref{eq:crh-sta-1}) we did not
make assumptions that the turbulence inhomogeneity scale is much larger
than the typical scale of turbulent flows, $\ell$. This assumption
is used in previous studies \citep{2010SoPh..tmp..241R}, and strictly
speaking their results can be applied only to the case of a weakly
stratified medium, $G\ell\ll1$.

\subsection{Calculation of $\overline{\mathbf{u}\cdot\mathbf{b}}$
in    the Fourier space}

{ Next, we estimate the contributions of terms in Eq.(\ref{eq:symmetry}) using
 the $\tau$-approximation made in
the Fourier space (see, e.g., \citealp{2000PhRvE..61.5202R,2003GApFD..97..249R}
and \citealp{2005PhR...417....1B}). Basically this approach has the same
shortcomings as Eq.(8). However, it allows  us to study the structure of
the cross-helicity in further detail. The derivations are explained
in Appendix A.}  In addition, we calculate correlation $\overline{u_{r}b_{r}}$,
which can be estimated from the line-of-sight observations of velocity
and magnetic field in a central part of the solar disk (cf., \citealp{2011SoPh..tmp...39Z}). 

Here we present the result of these calculations in the form that
can be compared with \citet{2003A&A...409.1097K} and \citet{2010SoPh..tmp..241R}:
\begin{eqnarray}
\overline{\mathbf{u}\cdot\mathbf{b}} & = & \frac{\eta_{T}}{2}\left(2+3\varepsilon\right)\left(\mathbf{\overline{B}}\cdot\mathbf{G}\right)\label{chr_f}\\
 & +\!\! & \!\!\frac{3\eta_{T}\left(\varepsilon+1\right)}{2}\left(\mathbf{\overline{B}}\cdot\mathbf{\mathbf{\nabla}}\right)\log\left(\overline{\mathbf{u}^{2}}\right)+\eta_{T}\tau_{c}\left(\varepsilon+1\right)\mu_{0}\left(\boldsymbol{\Omega}\cdot\overline{\mathbf{J}}\right)\nonumber \\
 & + & \frac{\eta_{T}\tau_{c}\mu_{0}}{4}\left(3+\varepsilon\right)\mathbf{\overline{J}}\cdot\mathbf{\overline{W}}+\frac{\eta_{T}\tau_{c}}{10}\left(23\varepsilon+5\right){\left(\boldsymbol{\nabla}\overline{\mathbf{U}}\right)_{\{i,j\}}}{\left(\boldsymbol{\nabla}\overline{\mathbf{B}}\right)_{\{i,j\}}},\nonumber 
\end{eqnarray}
where $\varepsilon$ is a ratio of the kinetic and magnetic energies
of the turbulent pulsations, $\eta_{T}=\overline{\mathbf{u}^{2}}\tau_{c}/3$
is turbulent diffusion is coefficient, and $\tau_{c}$ is a typical
convection turnover time. The structure of Eq.(\ref{chr_f})
corresponds to Eq.(\ref{eq:symmetry}) though contrary to Eq.(\ref{eq:crh-sta-1})
the contributions of density stratification ($\mathbf{G}$) and turbulent
intensity stratification ($\nabla\log\left(\overline{\mathbf{u}^{2}}\right)$)
are decoupled. For the other coefficients we find $\kappa_{3}=\mu_{0}\frac{\eta_{T}\tau_{c}}{4}\left(3+\varepsilon\right)$,
$\kappa_{4}=\mu_{0}\eta_{T}\tau_{c}\left(\varepsilon+1\right)$ and
$\kappa_{5}=\frac{\eta_{T}\tau_{c}}{10}\left(23\varepsilon+5\right)$.
Also, we see that the contribution of the stratification effects in
Eq.(\ref{eq:crh-sta-1}) agrees but have slightly different coefficients
$\frac{\left(2+3\varepsilon\right)}{2}$ vs $\left(1+\varepsilon\right)$,
and that the contributions due to the electric currents are in agreement.
Note, that here similarly to \citet{2010SoPh..tmp..241R} in derivation
of Eq.(\ref{chr_f}) we used the scale separation assumption, $\ell G\ll1$
(weakly stratified medium). 

For the correlation of the radial components we obtain: 
\begin{eqnarray}
\overline{u_{r}b_{r}} & = & \left(\overline{u_{r}b_{r}}\right)_{\rho}+\left(\overline{u_{r}b_{r}}\right)_{J\Omega}+\left(\overline{u_{r}b_{r}}\right)_{JW}\label{chr_r}\\
\left(\overline{u_{r}b_{r}}\right)_{\rho} & = & \frac{\eta_{T}G_{r}\overline{B}_{r}}{2}\left(2+\varepsilon\right)+\frac{\tau_{c}\left(\varepsilon+1\right)}{6}\overline{B}_{r}\nabla_{r}\overline{\mathbf{u}^{2}}\label{chr_rho}\\
 & - & \frac{\eta_{T}}{5}\left(5-2\varepsilon\right)\nabla_{r}\overline{B}_{r}\nonumber \\
\left(\overline{u_{r}b_{r}}\right)_{J\Omega} & = & \frac{2\varepsilon\eta_{T}\tau_{c}\mu_{0}}{5}\Omega_{r}J_{r}+\frac{4\varepsilon\eta_{T}\tau_{c}}{5}\Omega\cdot\left(\mathbf{e}^{(r)}\times\nabla_{r}\mathbf{\overline{B}}\right)\nonumber \\
\left(\overline{u_{r}b_{r}}\right)_{JW} & = & \frac{\eta_{T}\tau_{c}\mu_{0}}{10}\left(\varepsilon-3\right)\overline{J}_{r}\overline{W}_{r}+\frac{\eta_{T}\tau_{c}\mu_{0}}{20}\left(\varepsilon+7\right)\mathbf{\overline{J}}\cdot\mathbf{\overline{W}}\label{chr_JW}\\
 & + & \frac{\eta_{T}\tau_{c}}{70}\left(163\varepsilon+41\right)\nabla_{\{i}\overline{U}_{j\}}\nabla_{\{i}\overline{B}_{j\}},\nonumber 
\end{eqnarray}
where, $\mathbf{e}^{(r)}$ is a unit vector in radial direction. Both
Eq.(\ref{chr_f}) and Eq.(\ref{chr_r}) generalize the previous results
of \citet{2003A&A...409.1097K} and \citet{2010SoPh..tmp..241R} by
including the effects of the large-scale electric current and velocity
shear. In our models we find that the toroidal component of the large-scale
magnetic field is much stronger than the poloidal component. We also
discard the effect of meridional circulation in Eq.(\ref{chr_JW}).
Therefore we define the product of the magnetic and velocity strain
by formula:
\begin{eqnarray*}
\nabla_{\{i}\overline{U}_{j\}}\nabla_{\{i}\overline{B}_{j\}} & = & \nabla_{r}\overline{U}_{\phi}\nabla_{r}\overline{B}_{\phi}+\nabla_{\theta}\overline{U}_{\phi}\nabla_{\theta}\overline{B}_{\phi}+\nabla_{r}\overline{U}_{\phi}\nabla_{\theta}\overline{B}_{\phi}+\nabla_{\theta}\overline{U}_{\phi}\nabla_{r}\overline{B}_{\phi},
\end{eqnarray*}
where $\nabla_{r,\theta}$ are the covariant derivative components,
$\overline{B}_{\phi}$ is the toroidal magnetic field and $\overline{U}_{\phi}=r\sin\theta\left(\Omega\left(r,\theta\right)-\Omega_{0}\right)$
is the large-scale shear flow due to the differential rotation. 
\begin{figure}[ht]
\includegraphics[width=17cm,height=12cm]{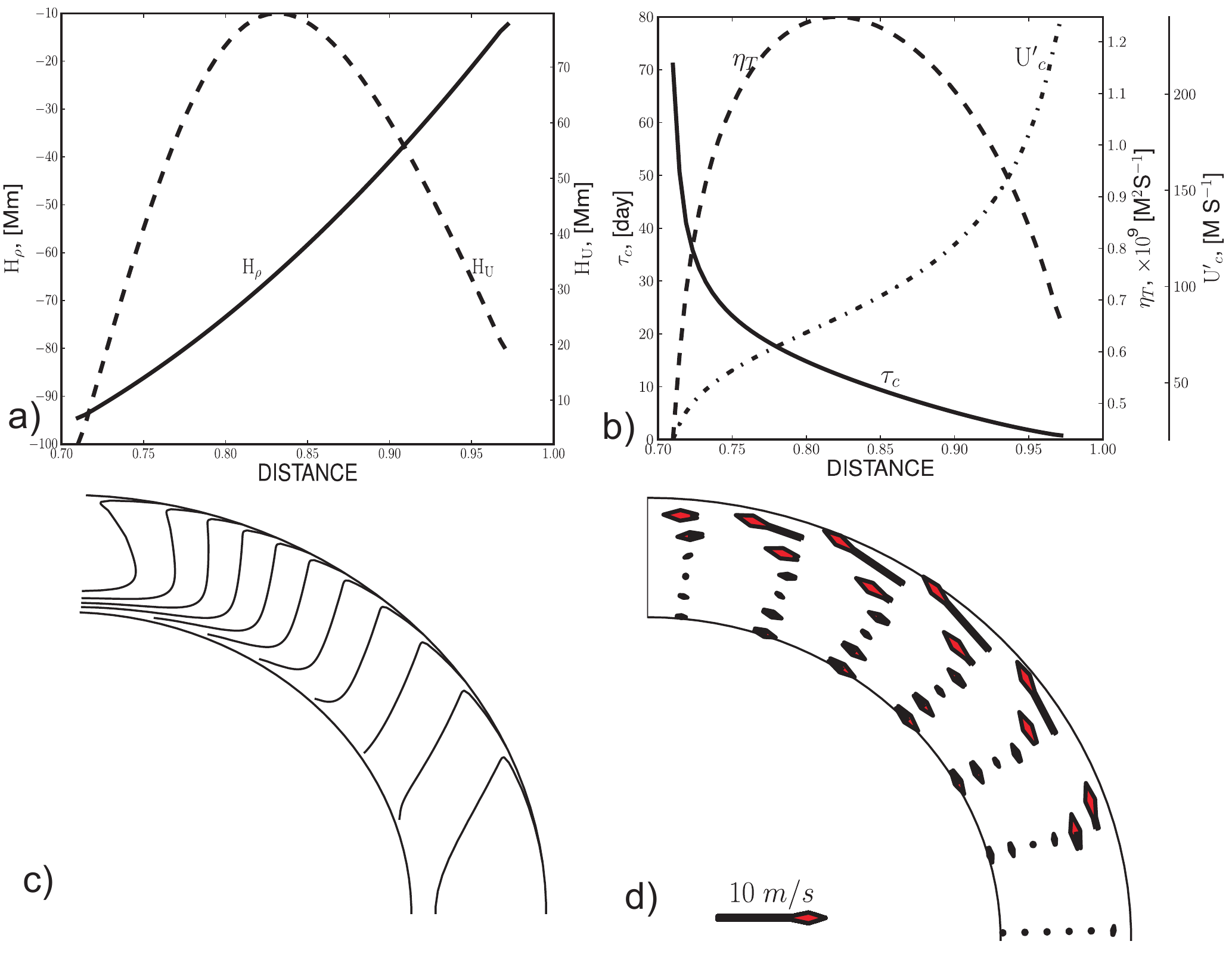}
\caption{Internal parameters of the solar convection zone: a), the density
scale $H_{\rho}=G^{-1}$ (solid line), and the turbulent velocity
scale $H_{U}={\displaystyle {(\nabla\log\bar{u^{2}})^{-1}}}$ (dashed
line), the distance is measured in units of the solar radius; b) the
typical turnover convective time, $\tau_{c}$ (solid line), the coefficient
of the turbulent diffusivity, $\eta_{T}$, (dashed line), the RMS
of convective velocity $u'$(dash-dot line); c) the contours of the
constant angular velocity are plotted for the levels \textbf{$(0.75-1.05)\Omega_{0}$};
d) the geometry of meridional circulation, the length of arrows are
proportional to the circulation speed.}

\label{Fig1} 
\end{figure}

\section{The  $\overline{u_{r}b_{r}}$ patterns by dynamo models}

In the next section, we illustrate contributions of each term to the
correlation of the radial components, $\overline{u_{r}b_{r}}$, given
by Eq.(\ref{chr_r}) for two types of solar dynamo models: the mean-field
dynamo distributed in the convection zone and the benchmark flux-transport
dynamo. The distributed dynamo model is described in detail by \citet{2009A&A...493..819P}
and \citet{2009A&A...508....9S}, hereafter, PS09 and SP09, respectively.
The benchmark model was constructed by \citet{2008A&A...483..949J}.
We consider the large-scale magnetic field produced by the models
as an input parameter for estimation of $\overline{u_{r}b_{r}}$ in
Eq.(\ref{chr_r}). The dynamo equations are given in Appendix B. 

The distributed dynamo model includes the meridional circulation with
the geometry of flow which is similar to that used by \citet{2002A&A...390..673B}.
The maximum velocity of the meridional circulation is fixed to 10~$\mathrm{ms^{-1}}$.
The turbulent generation effects include the $\alpha-$effect (Parker,
1955) and the ${\boldmath\mathbf{\Omega}}\times\mathbf{J}$ effect
\citep{rad69}. The internal parameters of the solar convection zone
are given by \citet{stix:02}. The integration domain is above the
tachocline, from 0.71$R_{\odot}$ to 0.972$R_{\odot}$ in radius,
and from the pole to the pole in latitude. We use the differential
rotation profile given by \citet{1998MNRAS.298..543A}. The internal
parameters of the model, including the turbulence parameters and parameters
of stratification of solar convection zone, distribution of the angular
velocity and the geometry of the meridional circulation are shown
in Figure~\ref{Fig1}.

\begin{figure*}[ht] 
\includegraphics[width=16cm,height=20cm]{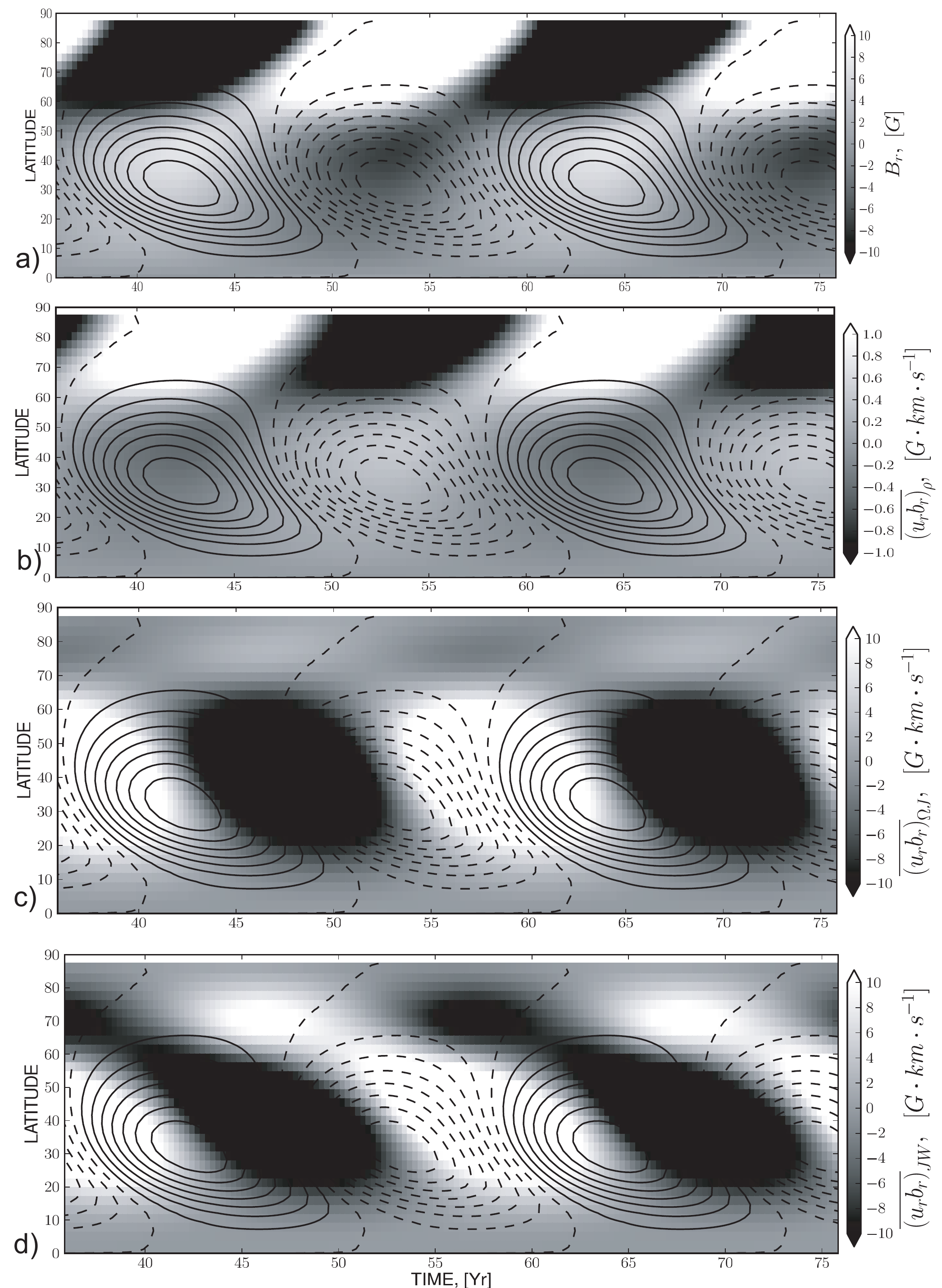}



 \caption{The time-latitude ({}``butterfly'') diagrams for the distributed
dynamo model. Contours in each plot the toroidal magnetic field at
the bottom of the convection zone. The color background shows: a)
the large-scale radial magnetic field at the surface; b) the cross-helicity
component $\left(\overline{u_{r}b_{r}}\right)_{\rho}$; c) $\left(\overline{u_{r}b_{r}}\right)_{\Omega J}$
; d) $\left(\overline{u_{r}b_{r}}\right)_{JW}$, defined in Eq. (\ref{chr_r},\ref{chr_rho}). }

\label{Fig2} 
\end{figure*}

\begin{figure*}[ht]
\includegraphics[width=16cm,height=20cm]{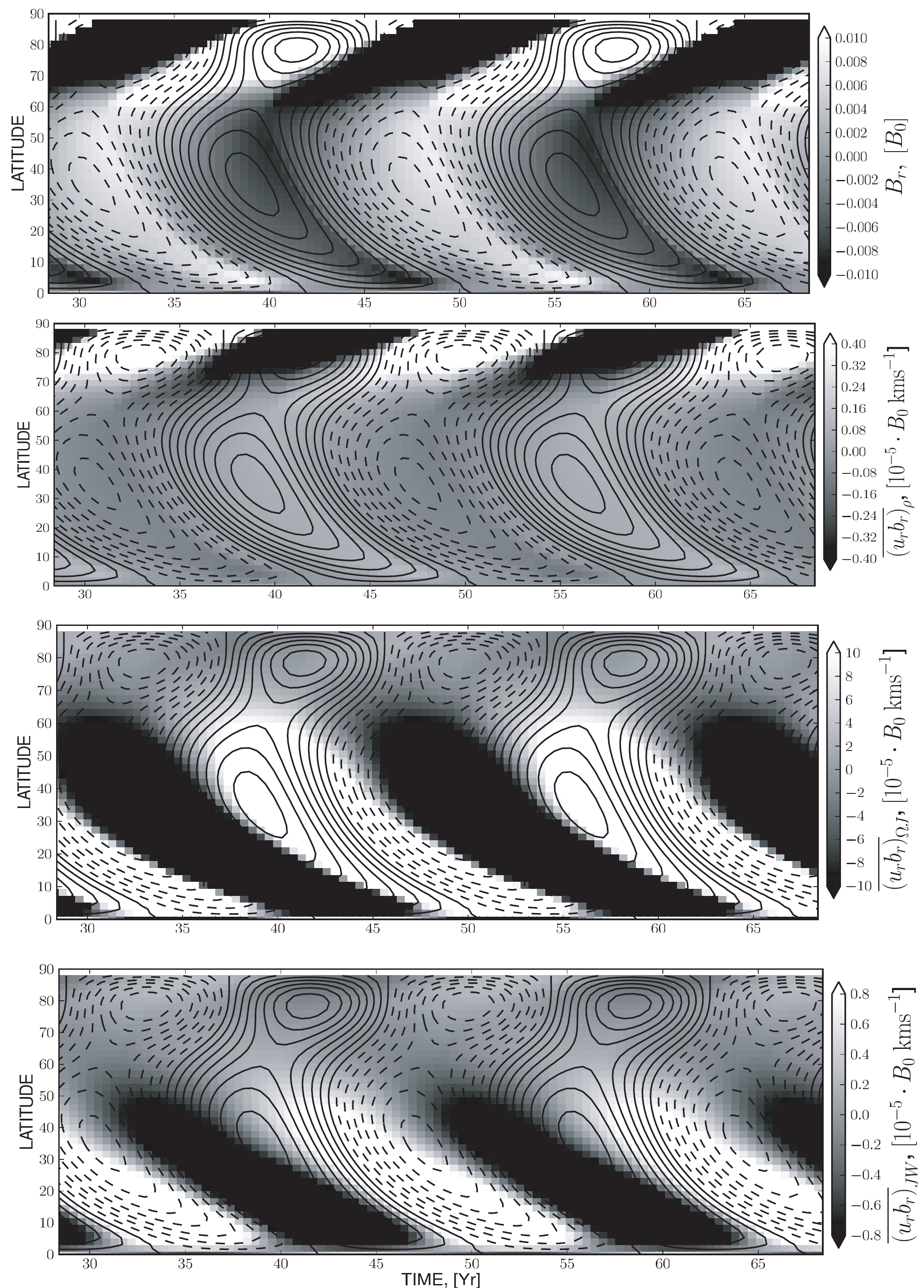}

\caption{The same as in Fig.\ref{Fig2} for the flux-transport dynamo model. }

\label{Fig3} 
\end{figure*}

For comparison we examine the $\overline{u_{r}b_{r}}$ patterns of
the benchmark flux-transport dynamo model. This type of dynamo models
is one of those widely discussed in the literature (e.g., \citealp{2002A&A...390..673B,2004ApJ...601.1136D}).
We keep the same formulation of the benchmark dynamo model as described
by \citet{2008A&A...483..949J}, although we use the same type of
rotation law as in the previous subsection. The integration domain in
this model is from $0.65R_{\odot}$ to $0.97R_{\odot}$.
The calculation of the $\overline{u_{r}b_{r}}$ is carried out from
$0.71R_{\odot}$ to $0.97R_{\odot}$. The magnetic field in this benchmark
model is measured in the units of $B_{0}$. Therefore the calculated
cross-helicity is scaled with the magnetic field strength $B_{0}$.
The other parameters of the model are: the reference turbulent diffusivity,
$5\cdot10^{10}\mathrm{cm^{2}s^{-1}}$ ($C_{\eta}=0.005$), the magnetic
Reynolds number, $\mathrm{Rm}=700$, the coefficient of the poloidal
magnetic field source term, $C_{s}=30$ (in notations of \citealp{2008A&A...483..949J}). 

We separate contributions to $\overline{u_{r}b_{r}}$ into three parts:
$\left(\overline{u_{r}b_{r}}\right)_{\rho}$ , $\left(\overline{u_{r}b_{r}}\right)_{\Omega J}$
and $\left(\overline{u_{r}b_{r}}\right)_{JW}$, and associate the
first term of Eq.(\ref{chr_r}) as a contribution of the stratification
effects, while the second and third terms represent contributions
of the large-scale electric current. In the calculations we assume
the equipartition between the kinetic and magnetic energies of fluctuations,
$\varepsilon=1$. Modeling the cross-helicity we found that if we
take the surface values for the model quantities in Eq.(\ref{chr_r})
then we obtain that the stratification effects are dominant. This
is true for the both types of dynamo models. However, if we integrate
the cross-helicity from the surface down to deeper layers, say down
to $0.9R_{\odot}$, then the contributions of the large-scale current
becomes comparable to the stratification effects and even greater.
The dynamo butterfly diagrams and the simulated patterns of the cross-helicity
variations are shown in Figures \ref{Fig2} and \ref{Fig3}. In the
simulations, the antisymmetric modes of the toroidal magnetic field
are dominant. Therefore we show the butterfly diagram only for the
northern hemisphere\textbf{.} The butterfly diagram of the toroidal
magnetic field is shown for the bottom of the convection zone (from
where the magnetic field of active regions presumably erupts), while
the radial magnetic field patterns are shown for the top of the integration
domain. In the first model the toroidal magnetic field strength varies
with the amplitude of $2.5$kG at the bottom of the convection zone.
The radial magnetic field has the maximum of the magnitude variations
at the poles, where it varies with an amplitude of $40$G at the surface.
The cross-helicity varies with amplitude $\pm10\ \mathrm{G\cdot km\cdot s^{-1}}$.
In the second model, the magnetic field strength is measured in the
units of $B_{0}$, which is assumed to be the maximum of the typical
strength of the flux tubes accumulated just beneath the solar convection
zone \citep{1999ApJ...518..508D}. The cross-helicity in the flux-transport
model is scaled with $B_{0}$ as well. In this model, $B_{0}$ should
be about $10^{5}$G in order to obtain the cross-helicity values similar
to the previous model. The main reason for such high $B_{0}$ value
is, of course, the low magnetic diffusivity used in the model. As
shown by \citet{2004ApJ...601.1136D}, the flux-transport models can
be tuned for high magnetic diffusivity in the near surface layers.
In both cases, one can see that the part of the cross-helicity, which
results from the density and turbulence stratification effects ( panel
(b) of the Figures 2 and 3), follows the evolution of the radial magnetic
fields while the large-scale current produces a more complicated pattern
of the cross-helicity of comparable magnitude, panels (c) and (d).
Both models give somewhat similar predictions for the cross-helicity
patterns. The most obvious difference is that the second model predicts
a very strong amplitude of the cross-helicity near the poles. This
effect results from the strong polar magnetic field which varies in
the range of $\pm0.3B_{0}$ while the toroidal magnetic field in this
model varies in the range of $\pm3B_{0}$. In Figure~3, in order
to resolve the fine structure of the sunspot formation zone we restrict
variations of $B_{r}$ within the limits of $\pm0.01B_{0}$. Comparing
with observations, the polar magnetic field seems to be too strong
in the benchmark flux-transport model. Perhaps this model needs further
tuning in a way similar to given by \citet{2004ApJ...601.1136D}.
Looking at the contribution of the cross-helicity produced by the
large-scale current we find that in the distributed dynamo model the
surface cross-helicity pattern correlates with the variations of $\bar{J_{r}}$,
and in the flux-transport model both components of the current are
important. The large-scale distribution of electric currents can be
deduced from synoptic magnetogram (\citealp{2000ApJ...528..999P,2009ScChG..52.1713W}).
For the current theory it would be interesting to compare the spatial-temporal
patterns of the cross-helicity and the large-scale current. These
patterns will evolve in phase if the contribution of the electric
current in Eq(\ref{chr_f}) dominates.

If we consider the observational problem of finding the $\overline{u_{r}b_{r}}$,
we have to choose suitable spatial and temporal scales for averaging
(in the spirit of \citealp{2010MNRAS.402L..30Z}), in order to distinguish
between the processes of the {}``local'' (noted in \citealp{2006PhRvL..97y5002M,2008PhRvL.100h5003M})
and {}``global'' alignments of velocity and magnetic fields. We
can take as a working hypothesis that for larger scales most contribution
to the observed cross-helicity comes from deeper layers of the solar
convective zone. This assumption is consistent with our approximations
but requires further investigations. In particular, our study neglects
turbulent fluxes of the cross-helicity. This assumes that the spatial
scale for averaging exceeds the typical scale of the turbulent velocity
variations.

\section{Discussion}

In this paper using the mean-field magnetohydrodynamics framework
we derive an equation for the cross-helicity evolution in the presence
of the large-scale magnetic fields and flows. This equation (see Eq.(\ref{eq:tur-cross-evol})) can be
used to construct the mean-field dynamo model based on the cross-helicity
effects (see, \citealp{2000ApJ...537.1039Y,2009MNRAS.399..273S}).
It is also can be used to estimate the cross-helicity distribution
in subsurface layer of the Sun. The structure of the cross-helicity
source terms in the evolution equation was studied on the base of
the MHD equations governing turbulent fluctuations of velocity and
magnetic fields in stratified media in the presence of the large-scale
magnetic fields and shear flow. We calculated the contributions of
various effects to the cross-helicity pseudo-scalar. For comparison
with solar observations we calculated the correlation of the radial
components of the turbulent velocity and magnetic field and estimated
these contributions for two different types of the solar dynamo.

Our results show that the turbulent cross-helicity reflects internal
properties of the dynamo processes and allow us to discriminate
among  different solar dynamo models. For analysis of the observational data involving
the cross-helicity it is important to consider averaging over small
and large scales. If cross-helicity data are averaged over a small
portion of the solar surface and short time periods, the results may
contain mainly the effect of the density stratification and the vertical
component of the large-scale magnetic field as shown by \citet{2010SoPh..tmp..241R}.
However, this type of averaging may meet some practical problems because
of the difficulty to distinguish between the processes of the global
and local alignment, of the turbulent velocity and magnetic fields.
For instance a similar kind of alignment for velocity and vorticity
is known as kinetic helicity (see \citealp{1985PhRvL..54.2505P}).
However, for different scales of convection motions near the surface
level one may observe different signs of the kinetic helicity at different
depths \citep{2004PhDT.......412Z}.

If the cross-helicity data are averaged over large spatial and temporal
scales, then other effects related to the solar dynamo may become
important (see Eq.(\ref{eq:tur-cross-evol})). In particular, the
cross-helicity patterns depend on the distribution of large-scale
electric currents and velocity shear. { In this case, the analysis becomes
more complicated, and the large-scale transport of cross-helicity
represented by
$\overline{\mathcal{F}}^{C}$ (see Eq.(\ref{eq:turb-stres0}))
has to be taken into account.} Furthermore, the cross-helicity equations
(\ref{chr_f}) and (\ref{chr_r}), may become invalid in the deeper
layers of the Sun due to the strong Coriolis force effects ($\Omega^{\star}\gg1$).

Our study is done in the anelastic approximation and does not take
into account the density fluctuations effects. However, it is known (e.g.,
\citealp{1993A&A...274..647K}) that in the turbulent coefficients
the density fluctuations effects comes with factor $g\bar{u^{2}}/C_{s}^{2}\approx\bar{u^{2}}/|H_{\rho}|$,
and may become important. Therefore, we anticipate further development
of the present theory regarding the turbulent transport and compressibility
effects. In this paper, we have shown that the spatially-temporal
patterns of cross-helicity depend on details of the dynamo mechanism.
This suggests to use the cross-helicity tracer as a diagnostic tool
for solar dynamo models. The forthcoming analysis of space and ground
based observations towards direct determination of the cross-helicity
challenges the future theoretical studies.

\paragraph{Acknowledgments.}
This work was funded by the National Natural Science Foundation
of China (NSFC), National Basic Research Program of China under grant 2000078401
and 2006CB806301, and Chinese Academy of Sciences under grant KLCX2-YW-T04
and the CAS Visiting Professorships for Senior International
Scientists, as well as joint Chinese-Russian collaborative project of
NNSF of China and RFBR of Russia under grant 08-02-92211, and also
RFBR grant 10-02-00148-a, 10-02-00960 and NASA LWS NNX09AJ85G grant.

\selectlanguage{american}%


\selectlanguage{english}%
\section*{Appendix A. Calculation of the cross-helicity in $\tau$-approximation in Fourier space}

Our approach for calculation of the cross-helicity is presented in
detail by \citet{pi08Gafd}, hereafter, P08. The important steps in
procedure were described earlier (see, e.g., \citealp{2000PhRvE..61.5202R,2003GApFD..97..249R}
and \citealp{2005PhR...417....1B}). The calculation is based on the equations
governing the evolution of the fluctuating velocity and magnetic fields
in a rotating coordinate system (see, also, \citealp{2000PhRvE..61.5202R,2003GApFD..97..249R,pi08Gafd}):
\begin{eqnarray}
\frac{\partial\mathbf{b}}{\partial t}\!\! & = & \!\!\!\!\nabla\!\times\!\left(\mathbf{u}\times\mathbf{\overline{B}}\!+\!\mathbf{\overline{U}}\times\mathbf{b}\right)\!+\!\eta\nabla^{2}\mathbf{b}\!+\!{\mathfrak{G}}\,,\label{induc1}\\
\frac{\partial m_{i}}{\partial t} & = & -2\left(\mathbf{\Omega}\times\mathbf{m}\right)_{i}-\nabla_{i}\left(p'-\frac{2}{3}\left(\mathbf{G\cdot m}\right)\nu+\frac{\left(\mathbf{b\cdot\overline{B}}\right)}{2\mu}\right)\nonumber \\
 & + & \nu\Delta m_{i}+\nu\left(\mathbf{G\cdot\nabla}\right)m_{i}+f_{i}+\mathfrak{F}_{i}\label{navie1}\\
 & + & \frac{1}{\mu_{0}}\nabla_{j}\left(\overline{B}_{j}b_{i}+\overline{B}_{i}b_{j}\right)-\nabla_{j}\left(\overline{U}_{j}m_{i}+\overline{U}_{i}m_{j}\right)\,.\nonumber 
\end{eqnarray}
They can be obtained from Eqs(\ref{eq:induc},\ref{eq:moment}) by
splitting velocity and magnetic fields into the mean and fluctuating
part (see Eq.(\ref{eq:decom})). Here, $\mathfrak{G}$, and $\mathfrak{F}$
stand for unspecified nonlinear contributions of the fluctuating fields,
${\mathbf{m}}=\bar{\rho}{\mathbf{u}}$, $\mathbf{G}=\nabla\log\bar{\rho}$
is the density stratification scale of the media, $p'$ the fluctuating
pressure component, $\mathbf{\Omega}$ is the angular velocity responsible
for the Coriolis force, $\mathbf{\overline{U}}$ is the mean flow
velocity, weakly varying in space, and $\mathbf{f}$ is a random force
driving the turbulence, $\eta$ and $\nu$ are the microscopic magnetic
diffusivity and viscosity, respectively. The further calculation is
done as follows. First, it is convenient to write equations (\ref{induc1})
and (\ref{navie1}) in the Fourier space: 
\begin{eqnarray}
\left(\frac{\partial}{\partial t}+\eta z^{2}\!\!\right)\hat{b}_{j}(\mathbf{z})\!\! & = &\!\! \mathrm{i}z_{l}\!\!\int\!\!\left[\widehat{m}_{j}(\mathbf{z-q)}\hat{\left(\frac{\overline{B}_{l}}{\rho}\right)}(\mathbf{q})-\widehat{m}_{l}(\mathbf{z-q)}\hat{\left(\frac{\overline{B}_{j}}{\rho}\right)}(\mathbf{q})\right]\mathrm{{d}}\mathbf{q}\label{induc2}\\
 & + & \mathrm{i}z_{l}\int\left[\widehat{b}_{l}(\mathbf{z-q)}\hat{\overline{V}}_{j}(\mathbf{q})-\widehat{b}_{j}(\mathbf{z-q)}\hat{\overline{V}}_{l}(\mathbf{q})\right]\mathrm{{d}}\mathbf{q}+\widehat{{\mathfrak{G}}}_{j}.\nonumber \\
\left(\frac{\partial}{\partial t}+\nu z^{2}+\mathrm{i}\nu\left(\mathbf{Gz}\right)\right)\hat{m}_{i}(\mathbf{z}) & = & \hat{f}_{i}+\hat{\mathfrak{F}}_{i}-2\left(\mathbf{\Omega}\hat{\mathbf{z}}\right)\left(\hat{\mathbf{z}}\times\hat{\mathbf{m}}\right)_{i}\label{navie2}\\
 & - & \mathrm{i}\pi_{if}(\mathbf{z)}z_{l}\int\left[\widehat{m}_{l}(\mathbf{z-q)}\hat{\overline{V}}_{f}(\mathbf{q})+\widehat{m}_{f}(\mathbf{z-q)}\hat{\overline{V}}_{l}(\mathbf{q})\right]\mathrm{{d}}\mathbf{q}\nonumber \\
 & + & \frac{\mathrm{i}}{\mu}\pi_{if}(\mathbf{z)}z_{l}\int\left[\widehat{b}_{l}(\mathbf{z-q)}\hat{\overline{B}}_{f}\left({\mathbf{q}}\right)+\widehat{b}_{f}(\mathbf{z-q)}\hat{\overline{B}}_{l}\left(q\right)\right]\mathrm{{d}}\mathbf{q},\nonumber 
\end{eqnarray}
where the turbulent pressure was excluded from (\ref{navie1}) by
convolution with tensor $\pi_{ij}(\mathbf{z)}=\delta_{ij}-\hat{z}_{i}\hat{z}_{j}$,
$\delta_{ij}$ is the Kronecker symbol and $\hat{\mathbf{z}}$ is
a unit wave vector. The equations for the second-order moments that
make contributions to the cross-helicity tensor can be found directly
from (\ref{induc2},\ref{navie2}). As the preliminary step we write
the equations for the second-order products of the fluctuating fields,
and make the ensemble averaging of them, 
\begin{eqnarray}
\frac{\partial}{\partial t}\left\langle \hat{m}_{i}\left(\mathbf{z}\right)\hat{b}_{j}\left(\mathbf{z}'\right)\right\rangle  & = & Th_{ij}^{\varkappa}(\mathbf{z},\mathbf{z'})-\left(\eta z'^{2}+\nu z^{2}+\mathrm{i}\nu\left(\mathbf{Gz}\right)\right)\left\langle \hat{m}_{i}\left(\mathbf{z}\right)\hat{b}_{j}\left(\mathbf{z}'\right)\right\rangle \label{eq:kappa1}\\
 &  & \mathrm{i}z'_{l}\int\left[\left\langle \hat{m}_{i}\left(\mathbf{z}\right)\hat{m}_{j}(\mathbf{z'-q)}\right\rangle \hat{\left(\frac{\overline{B}_{l}}{\rho}\right)}(\mathbf{q})-\right.\nonumber \\
 & - & \left.\left\langle \hat{m}_{i}\left(\mathbf{z}\right)\hat{m}_{l}(\mathbf{z'-q)}\right\rangle \hat{\left(\frac{\overline{B}_{j}}{\rho}\right)}(\mathbf{q})\right]\mathd\mathbf{q}-2\left(\mathbf{\Omega}\hat{\mathbf{z}}\right)\varepsilon_{iln}\hat{z_{l}}\left\langle \hat{m_{n}}(\mathbf{z})\hat{b}_{j}(\mathbf{z'})\right\rangle \nonumber \\
 & + & \mathrm{i}z'_{l}\int\left[\left\langle \hat{m}_{i}\left(\mathbf{z}\right)\widehat{b}_{l}(\mathbf{z'-q)}\right\rangle \hat{\overline{V}}_{j}(\mathbf{q})-\left\langle \hat{m}_{i}\left(\mathbf{z}\right)\widehat{b}_{j}(\mathbf{z'-q)}\right\rangle \hat{\overline{V}}_{l}(\mathbf{q})\right]\mathrm{{d}}\mathbf{q}\nonumber \\
 & - & \mathrm{i}\pi_{if}(\mathbf{z)}z_{l}\int\left[\left\langle \widehat{m}_{l}(\mathbf{z-q)}\hat{b}_{j}\left(\mathbf{z}'\right)\right\rangle \hat{\overline{V}}_{f}(\mathbf{q})+\left\langle \widehat{m}_{f}(\mathbf{z-q)}\hat{b}_{j}\left(\mathbf{z}'\right)\right\rangle \hat{\overline{V}}_{l}(\mathbf{q})\right]\mathrm{{d}}\mathbf{q}\nonumber \\
 & + & \frac{\mathrm{i}}{\mu}z_{l}\pi_{if}(\mathbf{z)}\int\left[\left\langle \widehat{b}_{l}(\mathbf{z-q)}\hat{b}_{j}\left(\mathbf{z}'\right)\right\rangle \overline{B}_{f}\left(\mathbf{q}\right)+\left\langle \widehat{b}_{f}(\mathbf{z-q)}\hat{b}_{j}\left(\mathbf{z}'\right)\right\rangle \overline{B}_{l}\left(q\right)\right]\mathd\mathbf{q},\nonumber \\
\frac{\partial}{\partial t}\left\langle \hat{m}_{i}\left(\mathbf{z}\right)\hat{m}_{j}\left(\mathbf{z}'\right)\right\rangle  & = & -2\left(\mathbf{\Omega}\hat{\mathbf{z}}\right)\varepsilon_{iln}\hat{z_{l}}\left\langle \hat{m_{n}}(\mathbf{z})\hat{m}_{j}(\mathbf{z'})\right\rangle -2\left(\mathbf{\Omega}\hat{\mathbf{z}}'\right)\varepsilon_{jln}\hat{z_{l}'}\left\langle \hat{m_{i}}(\mathbf{z})\hat{m}_{n}(\mathbf{z'})\right\rangle \label{secm1}\\
 & - & \mathrm{i}\pi_{if}(\mathbf{z)}z_{l}\int\left[\left\langle \widehat{m}_{l}(\mathbf{z-q)}\hat{m}_{j}\left(\mathbf{z}'\right)\right\rangle \hat{\overline{V}}_{f}(\mathbf{q})+\left\langle \widehat{m}_{f}(\mathbf{z-q)}\hat{m}_{j}\left(\mathbf{z}'\right)\right\rangle \hat{\overline{V}}_{l}(\mathbf{q})\right]\mathrm{{d}}\mathbf{q}\nonumber \\
 & - & \mathrm{i}\pi_{jf}(\mathbf{z')}z'_{l}\int\left[\left\langle \hat{m}_{i}\left(\mathbf{z}\right)\widehat{m}_{l}(\mathbf{z-q)}\right\rangle \hat{\overline{V}}_{f}(\mathbf{q})+\left\langle \hat{m}_{i}\left(\mathbf{z}\right)\widehat{m}_{f}(\mathbf{z-q)}\right\rangle \hat{\overline{V}}_{l}(\mathbf{q})\right]\mathrm{{d}}\mathbf{q}\nonumber \\
 & + & \frac{\mathrm{i}}{\mu}\pi_{if}\left(\mathbf{z}\right)z_{l}\int\left[\left\langle \widehat{b}_{l}(\mathbf{z-q)}\hat{m}_{j}\left(\mathbf{z}'\right)\right\rangle \hat{\overline{B}}_{f}\left({\mathbf{q}}\right)+\left\langle \widehat{b}_{f}(\mathbf{z-q)}\hat{m}_{j}\left(\mathbf{z}'\right)\right\rangle \hat{\overline{B}}_{l}\left(q\right)\right]\mathrm{{d}}\mathbf{q}\nonumber \\
 & + & \frac{\mathrm{i}}{\mu}\pi_{jf}(\mathbf{z')}z'_{l}\int\left[\left\langle \hat{m}_{i}\left(\mathbf{z}\right)\widehat{b}_{l}(\mathbf{z-q)}\right\rangle \hat{\overline{B}}_{f}\left({\mathbf{q}}\right)+\left\langle \hat{m}_{i}\left(\mathbf{z}\right)\widehat{b}_{f}(\mathbf{z-q)}\right\rangle \hat{\overline{B}}_{l}\left(q\right)\right]\mathrm{{d}}\mathbf{q}\nonumber \\
 & + & Th_{ij}^{v}(\mathbf{z},\mathbf{z'})-\nu\left(z'^{2}+z^{2}+i\left(\mathbf{Gz}\right)+i\left(\mathbf{Gz'}\right)\right)\left\langle \hat{m}_{i}\left(\mathbf{z}\right)\hat{m}_{j}\left(\mathbf{z}'\right)\right\rangle ,\nonumber \\
\frac{\partial}{\partial t}\left\langle \hat{b}_{i}\left(\mathbf{z}\right)\hat{b}_{j}\left(\mathbf{z}'\right)\right\rangle  & = & Th_{ij}^{h}(\mathbf{z},\mathbf{z'})-\left(\eta z'^{2}+\eta z^{2}\right)\left\langle \hat{b}_{i}\left(\mathbf{z}\right)\hat{b}_{j}\left(\mathbf{z}'\right)\right\rangle \label{eq:mag1}\\
 & + & \mathrm{i}z'_{l}\int\left[\left\langle \hat{b}_{i}\left(\mathbf{z}\right)\hat{m}_{j}(\mathbf{z'-q)}\right\rangle \hat{\left(\frac{\overline{B}_{l}}{\rho}\right)}(\mathbf{q})-\left\langle \hat{b}_{i}\left(\mathbf{z}\right)\hat{m}_{l}(\mathbf{z'-q)}\right\rangle \hat{\left(\frac{\overline{B}_{j}}{\rho}\right)}(\mathbf{q})\right]\mathd\mathbf{q}\nonumber \\
 & + & \mathrm{i}z_{l}\int\left[\left\langle \hat{m}_{i}(\mathbf{z-q)}\hat{b}_{j}\left(\mathbf{z}'\right)\right\rangle \hat{\left(\frac{\overline{B}_{l}}{\rho}\right)}(\mathbf{q})-\left\langle \hat{m}_{l}(\mathbf{z-q)}\hat{b}_{j}\left(\mathbf{z}'\right)\right\rangle \hat{\left(\frac{\overline{B}_{i}}{\rho}\right)}(\mathbf{q})\right]\mathd\mathbf{q}\nonumber \\
 & + & \mathrm{i}z'_{l}\int\left[\left\langle \hat{b}_{i}\left(\mathbf{z}\right)\widehat{b}_{l}(\mathbf{z'-q)}\right\rangle \hat{\overline{V}}_{j}(\mathbf{q})-\left\langle \hat{b}_{i}\left(\mathbf{z}\right)\widehat{b}_{j}(\mathbf{z'-q)}\right\rangle \hat{\overline{V}}_{l}(\mathbf{q})\right]\mathrm{{d}}\mathbf{q},\nonumber 
\end{eqnarray}
where, the terms $Th_{ij}^{(\varkappa,v,h)}$ involve the third-order
moments of fluctuating fields and second-order moments of them with
the forcing term. To proceed further we introduce the double Fourier
transformation of an ensemble average of two fluctuating quantities,
say $f$ and $g$, taken at identical times and at the different positions
$\mathbf{x},\,\mathbf{x}'$, is given by 
\begin{equation}
\overline{f\left(\mathbf{x}\right)g\left(\mathbf{x}'\right)}\!=\!\int\int\overline{\hat{f}\left(\mathbf{z}\right)\hat{g}\left(\mathbf{z}'\right)}e^{\mathrm{i}\left(\mathbf{z}\cdot\mathbf{x}+\mathbf{z}'\cdot\mathbf{x}'\right)}\mathd^{3}\mathbf{z}\mathd^{3}\mathbf{z}'.\label{cor1}
\end{equation}
In the spirit of the general formalism of the two-scale approximation
(\citealp{rob-saw}) we introduce {}``fast'' and {}``slow'' variables.
They are defined by the relative $\mathbf{r}=\mathbf{x}-\mathbf{x}'$
and mean $\mathbf{R}=\left(\mathbf{x}+\mathbf{x}'\right)/2$, coordinates
respectively. The fast and slow variables in the Fourier space correspond
to two wave vectors: $\mathbf{k}=\left(\mathbf{z}-\mathbf{z}'\right)/2$
and $\mathbf{K}=\mathbf{z}+\mathbf{z}'$. Then, we define a correlation
function of $\hat{\mathbf{f}}$ and $\hat{\mathbf{g}}$ obtained from
Eq.(\ref{cor1}) by integration with respect to $\mathbf{K}$, 
\begin{equation}
\Phi\left(\hat{f},\hat{g},\mathbf{k},\mathbf{R}\right)=\int\overline{\hat{f}\left(\mathbf{k}+\frac{\mathbf{K}}{2}\right)\hat{g}\left(\frac{\mathbf{K}}{2}-\mathbf{k}\right)}\mathe^{\mathrm{i}\left(\mathbf{K}\cdot\mathbf{R}\right)}\mathd^{3}\mathbf{K}.\label{eq:cor1a}
\end{equation}
 For further convenience we define the following second-order correlations
of the momentum density, magnetic fluctuations, and cross-correlations
of the momentum and magnetic fluctuations: 
\begin{eqnarray}
\hat{v}_{ij}\left(\mathbf{k},\mathbf{R}\right)\!\! & \!=\! & \!\!\Phi(\hat{m}_{i},\hat{m}_{j},\mathbf{k},\mathbf{R}),\nonumber \\
\bar{\rho}^{2}\bar{\mathbf{u}^{2}}\left(\mathbf{R}\right)\! & = & \!\int\!\!\hat{v}_{ii}\left(\mathbf{k},\mathbf{R}\right)\mathd^{3}\mathbf{k},\nonumber \\
\hat{h}_{ij}\left(\mathbf{k},\mathbf{R}\right)\!\! & \!=\! & \!\!\Phi(\hat{b}_{i},\hat{b}_{j},\mathbf{k},\mathbf{R}),\nonumber \\
\bar{\mathbf{b}^{2}}\left(\mathbf{R}\right)\!\! & = & \!\!\int\!\!\hat{h}_{ii}\left(\mathbf{k},\mathbf{R}\right)\mathd^{3}\mathbf{k},\\
\hat{\varkappa}_{ij}\left(\mathbf{k},\mathbf{R}\right) & = & \Phi(\hat{m}_{i},\hat{b}_{j},\mathbf{k},\mathbf{R}),\nonumber 
\end{eqnarray}
We now return to equations (\ref{eq:kappa1}), (\ref{secm1}) and
(\ref{eq:mag1}). As the first step, we approximate the $Th_{ij}^{(\varkappa,v,h)}$-terms
by the corresponding $\tau$ relaxation terms of the second-order
contributions,

\begin{eqnarray}
Th_{ij}^{(\varkappa)} & \approx & -\frac{\left\langle \hat{m}_{i}\left(\mathbf{z}\right)\hat{b}_{j}\left(\mathbf{z}'\right)\right\rangle }{\tau_{c}},\label{eq:thx}\\
Th_{ij}^{(v)} & \approx & -\frac{\left\langle \hat{m}_{i}\left(\mathbf{z}\right)\hat{m}_{j}\left(\mathbf{z}'\right)\right\rangle -\left\langle \hat{m}_{i}\left(\mathbf{z}\right)\hat{m}_{j}\left(\mathbf{z}'\right)\right\rangle ^{\left(0\right)}}{\tau_{c}},\label{eq:thv}\\
Th_{ij}^{(h)} & \approx & -\frac{\left\langle \hat{b}_{i}\left(\mathbf{z}\right)\hat{b}_{j}\left(\mathbf{z}'\right)\right\rangle -\left\langle \hat{b}_{i}\left(\mathbf{z}\right)\hat{b}_{j}\left(\mathbf{z}'\right)\right\rangle ^{(0)}}{\tau_{c}},\label{eq:thh}
\end{eqnarray}
where the superscript $^{\left(0\right)}$ denotes the moments of
the background turbulence. Here, $\tau_{c}$is independent on $\mathbf{k}$
and it is independent on the mean fields as well.
 As the next step
we make the Taylor expansion with respect to the {}``slow'' variables
and take the Fourier transformation, (\ref{eq:cor1a}), about them.
The details of this procedure can be found in \citep{2005PhR...417....1B}.
Furthermore, we consider the high Reynolds number limit (relevant
for the solar turbulence) and discard the microscopic diffusion terms.
Then we  arrive to the evolution equation for the cross-helicity tensor
(see, P08): 
\begin{eqnarray}
\frac{\partial\hat{\varkappa}_{ij}}{\partial t} & = & -\mathrm{i}\left(\mathbf{\overline{B}k}\right)\left(\frac{\hat{v}_{ij}}{\bar{\rho}}-\frac{\hat{h}_{ij}}{\mu_{0}}\right)+\frac{\left(\overline{\mathbf{B}}\nabla\right)}{2}\left(\frac{\hat{v}_{ij}}{\bar{\rho}}+\frac{\hat{h}_{ij}}{\mu_{0}}\right)\label{eq:secm2a}\\
 & + & \frac{\left(\overline{\mathbf{B}}\mathbf{k}\right)}{2\bar{\rho}}G_{s}\frac{\partial\hat{v}_{ij}}{\partial k_{s}}-\frac{\left(\mathbf{G\overline{B}}\right)}{2\bar{\rho}}\hat{v}_{ij}-\frac{1}{\mu_{0}}\hat{k}_{i}\hat{k}_{f}\overline{B}_{f,l}\hat{h}_{lj}\nonumber \\
 & + & \frac{1}{\bar{\rho}}G_{l}\hat{v}_{il}B_{j}+\frac{\hat{h}_{lj}\overline{B}_{i,l}}{\mu_{0}}-\frac{\hat{v}_{il}\overline{B}_{j,l}}{\bar{\rho}}-\frac{\hat{\varkappa}_{ij}}{\tau_{c}}+\overline{U}_{j,l}\hat{\varkappa}_{il}\nonumber \\
 & - & \frac{k_{l}\overline{B}_{l,f}}{2}\frac{\partial}{\partial k_{f}}\left[\frac{\hat{v}_{ij}}{\bar{\rho}}+\frac{\hat{h}_{ij}}{\mu_{0}}\right]-\overline{U}_{i,l}\hat{\varkappa}_{lj}+\left[\hat{k}_{i}\hat{k}_{f}\right.\nonumber \\
 & - & \left.-\frac{\mathrm{i}}{2k}\left(2\hat{k}_{i}\hat{k}_{f}\left(\mathbf{\hat{k}}\cdot\mathbf{\nabla}\right)-\hat{k}_{i}\nabla_{f}-\hat{k}_{f}\nabla_{i}\right)\right]\hat{\varkappa}_{lj}\overline{U}_{f,l}\nonumber \\
 & - & 2\left(\Omega\hat{k}\right)\hat{k}_{p}\varepsilon_{ipl}\hat{\varkappa}_{lj}-2\frac{\mathrm{i}}{k}\left(\Omega\hat{k}\right)\hat{k}_{p}\varepsilon_{ipl}\left(\hat{k}\nabla\right)\hat{\varkappa}_{lj}\nonumber \\
 & + & \frac{\mathrm{i}}{k}\varepsilon_{ipl}\left(\left(\Omega\hat{k}\right)\nabla_{p}\hat{\varkappa}_{lj}+\hat{k}_{p}\left(\Omega\nabla\right)\hat{\varkappa}_{lj}\right)+k_{l}\overline{U}_{f,l}\frac{\partial\hat{\varkappa}_{ij}}{\partial k_{f}},\nonumber 
\end{eqnarray}
where $\hat{\mathbf{k}}$ is a unit wave vector, $\overline{U}_{f,l}\equiv\nabla_{l}\overline{U}_{f}$
and $\overline{B}_{f,l}\equiv\nabla_{l}\overline{B}_{f}$. The expressions
for the Fourier image of cross-helicity pseudo-scalar can be obtained
from Eq.(\ref{eq:secm2a}) if we assume quasi-stationarity (discarding
the time-derivative terms). Then we solve the obtained equation in
the linear approximation with respect to the large-scale fields and
the Coriolis force and obtain the cross-helicity tensor $\hat{\varkappa}_{ij}$.
After this we calculate the cross-correlations:
\begin{eqnarray}
\overline{\mathbf{u}\cdot\mathbf{b}} & = & \frac{1}{\overline{\rho}}\int\!\!\hat{\varkappa}_{ii}\left(\mathbf{k},\mathbf{R}\right)\mathd^{3}\mathbf{k},\\
\overline{u_{r}b_{r}} & = & \frac{1}{\overline{\rho}}\int\!\! e_{i}^{(r)}e_{j}^{(r)}\hat{\varkappa}_{ij}\left(\mathbf{k},\mathbf{R}\right)\mathd^{3}\mathbf{k},
\end{eqnarray}
where $e_{i}^{(r)}$ is the unit vector in radial direction. For the
final results we have to define the spectra of the background turbulence.
We consider the case of stationary and quasi-isotropic background
turbulence\citep{1987GApFD..38..273K},

\begin{eqnarray}
\hat{v}_{ij} & = & \left\{ \pi_{ij}\left(\mathbf{k}\right)+\frac{\mathrm{i}}{2k^{2}}\left(k_{i}\nabla_{j}-k_{j}\nabla_{i}\right)\right\} \frac{\bar{\rho}^{2}E\left(k,\mathbf{R}\right)}{8\pi k^{2}},\label{eq:spectr1}
\end{eqnarray}
 where $\pi_{ij}(\mathbf{k)}=\delta_{ij}-\hat{k}_{i}\hat{k}_{j}$,
$\delta_{ij}$ is the Kronecker symbol, the spectral function $E(k,\mathbf{R})$
defines the intensity of the velocity fluctuations: 
\begin{eqnarray}
\overline{\mathbf{u}^{2}} & = & \int\frac{E\left(k,\mathbf{R}\right)}{4\pi k^{2}}\mathd^{3}\mathbf{k}.\label{eq:spectr3-1u}
\end{eqnarray}
Similarly for the magnetic fluctuations of the turbulence, by the
spectral function $\mathcal{B}(k,\mathbf{R})$ : 
\begin{eqnarray}
\overline{\mathbf{b}^{2}} & = & \int\frac{\mathcal{B}\left(k,\mathbf{R}\right)}{4\pi k^{2}}\mathd^{3}\mathbf{k}.\label{eq:spectr3}
\end{eqnarray}
The anisotropy in Eq.(\ref{eq:spectr1}) is induced by inhomogeneity
of the turbulence. However, the scale of inhomogeneity is assumed
to be small compared to the typical scale of turbulent flows. { Therefore,
the turbulent anisotropy is small and the given
model of the background turbulence is called as quasi-isotropic }\citep{1987GApFD..38..273K}.
We omit the further cumbersome derivations, which are explained in
detail in P08.

\section*{Appendix B. Dynamo model equations}

The evolution of the axisymmetric magnetic field ($B$ being the azimuthal
component of the magnetic field, $A$ is proportional to the azimuthal
component of the vector potential) is governed by the following equations:
\begin{eqnarray}
\frac{\partial A}{\partial t}\!\! & \!\!=\!\! & \!\! r\sin\theta\mathcal{E}_{\phi}+\frac{U_{\theta}\sin\theta}{r}\frac{\partial A}{\partial\mu}-U_{r}\frac{\partial A}{\partial r}\label{eq:A}\\
r\sin\theta\mathcal{E}_{\phi}\!\! & \!\!=\!\! & \!\!\phi_{6}^{(a)}C_{\alpha}\eta_{T}Gr\mu\sin\theta Bf_{12}^{(a)}\label{eq:e_phi}\\
 & + & \psi_{\eta}\eta_{T}\left\{ f_{2}^{(d)}+2f_{1}^{(a)}\right\} \frac{\partial^{2}A}{\partial r^{2}}\nonumber \\
 & + & \psi_{\eta}\eta_{T}\left\{ f_{2}^{(d)}+2f_{1}^{(a)}\right\} \frac{\left(1-\mu^{2}\right)}{r^{2}}\frac{\partial^{2}A}{\partial\mu^{2}}\nonumber \\
 & + & \psi_{\delta}C_{\delta}\eta_{T}f_{4}^{(d)}\sin\theta\left(r\mu\frac{\partial B}{\partial r}+\left(1-\mu^{2}\right)\frac{\partial B}{\partial\mu}\right)\nonumber 
\end{eqnarray}
 
\begin{eqnarray}
\frac{\partial B}{\partial t} & = & -\sin\theta\left(\frac{\partial\Omega}{\partial r}\frac{\partial A}{\partial\mu}-\frac{\partial\Omega}{\partial\mu}\frac{\partial A}{\partial r}\right)-\frac{\partial\left(rU_{r}B\right)}{\partial r}\label{eq:B}\\
 & + & \frac{\sin\theta}{r}\frac{\partial U_{\theta}B}{\partial\mu}+\frac{1}{r}\frac{\partial r\mathcal{E}_{\theta}}{\partial r}+\frac{\sin\theta}{r}\frac{\partial\mathcal{E}_{r}}{\partial\mu}\nonumber \\
\frac{\sin\theta}{r}\frac{\partial\mathcal{E}_{r}}{\partial\mu} & = & \frac{\sin\theta}{r^{2}}\frac{\partial}{\partial\mu}\left(\eta_{T}\psi_{\eta}\left(f_{2}^{(d)}\right.\right.\nonumber \\
 & + & \left.\left.2f_{1}^{(a)}\left(1-\mu^{2}\right)\right)\right)\frac{\partial\sin\theta B}{\partial\mu}\label{eq:e_r}\\
 & + & \frac{2f_{1}^{(a)}\sin\theta}{r^{2}}\frac{\partial}{\partial r}\psi_{\eta}\frac{\partial}{\partial\mu}\left(\mu\sin\theta B\right)+\nonumber \\
 & + & \frac{f_{1}^{(a)}\sin\theta}{r}G\frac{\partial}{\partial\mu}\left(\psi_{\eta}\mu\sin\theta B\right)\nonumber \\
\frac{1}{r}\frac{\partial r\mathcal{E}_{\theta}}{\partial r} & = & \frac{1}{r}\frac{\partial}{\partial r}\left(\eta_{T}\psi_{\eta}\left(f_{2}^{(d)}+2f_{1}^{(a)}\mu^{2}\right)\right)\frac{\partial rB}{\partial r}+\label{eq:e_th}\\
 & + & \frac{2\mu\sin\theta}{r}\frac{\partial}{\partial r}\eta_{T}f_{1}^{(a)}\psi_{\eta}\frac{\partial}{\partial\mu}\left(\sin\theta B\right)\nonumber \\
 & - & \frac{1}{r}\frac{\partial}{\partial r}r\psi_{\eta}\left(Gf_{3}^{(a)}+G(2\mu^{2}-1)f_{1}^{(a)}\right)B\nonumber 
\end{eqnarray}
 where $\mu=\cos\theta$, $\theta$ is co-latitude. We use the same
notations as in P08. Functions $f_{1,2}^{(a,d)}$ depend on the Coriolis
number $\Omega^{*}=2\tau_{c}\Omega_{0}$, $\tau_{c}$ is a typical
correlation time of the turbulent convection; functions $\phi_{6}^{(a)}$
and $\psi_{\eta,\delta}$ describe magnetic quenching and depend on
$\beta=B/\sqrt{\mu_{0}\bar{\rho}\bar{u^{2}}}$, parameters $C_{\alpha},\, C_{\delta}$
control the contributions of the $\alpha$ and $\Omega\times J$ effects.
In notations of P08, $\psi_{\eta}=\phi_{3}+\phi_{2}-2\phi_{1}$ and
$\psi_{\delta}=\phi_{7}^{(w)}+\phi_{2}^{(w)}$. We introduce parameter
$C_{\eta}$ to control the turbulent diffusion coefficient, $\eta_{T}=C_{\eta}\tau_{c}\bar{\mathbf{u}^{2}}/3$.
Following to the linear analysis of PS09 and SP09 we choose $C_{\delta}=0.02,C_{\alpha}=C_{\delta}/2,C_{\eta}=0.1$.
The dynamo equations with these input parameters give a solar-type
dynamo solution. 
\end{document}